\documentclass[conference]{IEEEtran}
\IEEEoverridecommandlockouts
\usepackage{cite}
\usepackage{amsmath,amssymb,amsfonts}
\usepackage{algorithmic}
\usepackage{graphicx}
\usepackage{textcomp}
\usepackage{xcolor}
\definecolor{darkgreen}{RGB}{0,128,0}
\definecolor{darkred}{RGB}{178,34,34}
\usepackage{pifont}
\usepackage{multirow}
\usepackage{graphicx}
\newcommand{\cmark}{\ding{51}}  % check mark
\newcommand{\xmark}{\ding{55}}  % cross mark
\usepackage{subcaption}
\usepackage{url}  % Allows proper URL formatting
\usepackage[breaklinks=true]{hyperref}  % Enables link breaking
\usepackage{caption}
\def\BibTeX{{\rm B\kern-.05em{\sc i\kern-.025em b}\kern-.08em
    T\kern-.1667em\lower.7ex\hbox{E}\kern-.125emX}}
\begin{document}

\title{Data-Efficient Point Cloud Semantic Segmentation Pipeline for Unimproved Roads}

\author{\IEEEauthorblockN{Andrew Yarovoi}
\IEEEauthorblockA{\textit{Woodruff School of Mechanical
 Engineering} \\
\textit{Georgia Institute of Technology}\\
Atlanta, USA \\
ayarovoi3@gatech.edu}
\and
\IEEEauthorblockN{Christopher R. Valenta}
\IEEEauthorblockA{\textit{Electro-Optical Systems Laboratory} \\
\textit{Georgia Tech Research Institute}\\
Atlanta, USA \\
chris.valenta@gtri.gatech.edu}
}

\maketitle

\begin{abstract}
In this case study, we present a data-efficient point cloud segmentation pipeline and training framework for robust segmentation of unimproved roads and seven other classes. Our method employs a two-stage training framework: first, a projection-based convolutional neural network is pre-trained on a mixture of public urban datasets and a small, curated in-domain dataset; then, a lightweight prediction head is fine-tuned exclusively on in-domain data. Along the way, we explore the application of Point Prompt Training to batch normalization layers and the effects of Manifold Mixup as a regularizer within our pipeline. We also explore the effects of incorporating histogram-normalized ambients to further boost performance. Using only 50 labeled point clouds from our target domain, we show that our proposed training approach improves mean Intersection-over-Union from 33.5\% to 51.8\% and the overall accuracy from 85.5\% to 90.8\%, when compared to naive training on the in-domain data. Crucially, our results demonstrate that pre-training across multiple datasets is key to improving generalization and enabling robust segmentation under limited in-domain supervision. Overall, this study demonstrates a practical framework for robust 3D semantic segmentation in challenging, low-data scenarios. Our code is available at: \url{https://github.com/andrewyarovoi/MD-FRNet}.
\end{abstract}

\section{Introduction}
Semantic segmentation of 3D point clouds is a foundational task for scene understanding, enabling a range of downstream applications such as autonomous route planning and infrastructure inspection. Despite significant progress in this field, most state-of-the-art segmentation models rely heavily on the availability of large, labeled training datasets. However, generating labeled point cloud data remains a substantial bottleneck: manual annotation is both labor-intensive and time-consuming, requiring over 30 minutes per scan on average in our experiments. This challenge makes it impractical to recreate large-scale datasets, commonly containing over 25,000 scans, for new or underrepresented environments. 

In this case study, we explore training strategies and model adaptations designed to improve data efficiency for 3D semantic segmentation, with the goal of achieving robust performance using only 50 labeled point clouds. Our target domain consists of rural, dirt, and gravel roads, forest trails, and paved paths in open grassland environments. This setting stands in stark contrast to widely-used public datasets such as the Waymo Open Dataset \cite{Sun2020ScalabilityDataset} and SemanticKITTI \cite{Behley2019SemanticKITTI:Sequences}, which primarily focus on urban driving scenes featuring paved roads, buildings, and sparse vegetation. Moreover, our point clouds are captured using an Ouster OS-1 LiDAR sensor, which differs significantly from the sensors used in those benchmarks, particularly in vertical field-of-view (FOV) and angular resolution. These domain and sensor disparities make direct application of models trained on public datasets ineffective for the target application. 

To address this challenge, we draw on insights from prior work in few-shot learning, transfer learning, and multi-dataset synergistic training. Our main contributions are as follows:

\begin{itemize} 
    \item We propose a practical and effective framework for training point cloud semantic segmentation models using limited in-domain data supplemented by publicly available out-of-domain datasets. 
    \item We evaluate the effectiveness of Point Prompt Training (PPT) when applied to convolutional architectures with batch normalization, expanding its application beyond transformer-based models. 
    \item We investigate the use of Manifold Mixup (MM) for regularizing point cloud segmentation heads and improving generalization, extending its application beyond prior work focused on image classification and data augmentation. 
\end{itemize}

In the following sections, we first present a survey of relevant literature in section II. We then present our methodology in section III and our results in section IV. Finally, we provide a discussion of our work and present future plans in section V.

\section{Literature Review}
Given the limited number of training samples in our target dataset, this study can be framed as an application of Few-Shot Learning (FSL). While FSL has been widely explored in the context of image classification, its application to segmentation and 3D point clouds has received comparatively less attention. In the following review, we discuss common FSL strategies across both image and point cloud modalities, and introduce Point Prompt Training and Manifold Mixup as they relate to our proposed methodology.

\subsection{Few-Shot Learning on Images}
Few-shot learning (FSL) has been extensively studied for image-based tasks. The goal of FSL is to enable models to quickly learn to classify samples drawn from new classes (the query set) with minimal labeled examples of the new classes (support set) by leveraging prior knowledge, learned feature representations, and efficient adaptation mechanisms \cite{Chen2019AClassification}. In the standard FSL setting, the test set and the training set contain a large domain shift, typically enforced by ensuring that the two sets contain entirely disjoint classes. During training, a large base dataset is provided and labels are available for all samples. During testing, a small dataset (support set) containing only a few labeled examples of the test classes (not found in the training set) is provided, and the network must predict the labels for the remaining samples (query set). In our case study, large-scale urban datasets serve as the base training data, while our labeled in-domain training dataset functions as the support set. The labeled in-domain test samples are then used as the query set to evaluate model performance. Unlike standard FSL, we do not impose restrictions on using the support set during initial training, as our sole objective is to optimize performance on the testing query set.

Two of the most prominent approaches for FSL on images are metric-based learning and transfer learning. In metric-based learning, the base dataset is used to train an embedding model to map semantically similar classes to nearby feature embeddings. During testing, query embeddings are assigned the labels of their nearest support embeddings based on a chosen distance metric.

One of the earliest and most influential metric-based models was Prototypical Network (ProtoNet) \cite{Snell2017PrototypicalLearning}. ProtoNet trained an embedding network on a diverse base dataset and represented the novel, unseen classes within the support set using prototype vectors, computed as the mean of the support set embeddings. Query samples were then classified based on their Euclidean distance to these prototypes within the embedding space. This approach enabled rapid generalization to new classes with only a few labeled support examples (typically 1 to 5), making it a foundational method in few-shot learning.

Following ProtoNet, lots of works explored varied metric-based assignment schemes for associating query images to a small number of provided support samples. SimpleShot \cite{Wang2019SimpleShot:Learning} assigned query embeddings to classes based on their nearest neighbor within the support embeddings. \cite{Chen2019Self-SupervisedClassification} pretrained the embedding network on the training set (and optionally other external datasets) using a self-supervised vision task. Then it fine-tuned a prototypical network on the supervised training data. When pretraining on larger external datasets, the model showed significant improvements in classification accuracy. \cite{Hu2022PushingDifference} also explored self-supervised training of a prototypical network and compared it to utilizing supervised pretraining on external datasets.

While metric-based learning is the most prominent approach to FSL, transfer learning has also demonstrated competitive performance. In transfer learning, an embedding network is trained on the base training set to embed individual samples into consistent feature embeddings (similar to metric-based models). However, instead of using proximity to support feature embeddings to label query features during test time, the feature embedding model is frozen and a small classifier is trained to map the feature embeddings directly to the novel classes using supervision from the small support set. To prevent overfitting on such a small support set, the classifier model must be very small or be heavily regularized. For example, \cite{Tian2020RethinkingNeed} used a simple linear classifer trained on the extremely limited support set to directly assign class labels to the query set. \cite{Chen2019AClassification} explored various simple classifiers trained on the support set and investigated larger embedding networks, which were found to provide significant improvements in generalization to new classes. Both models produced competitive results to metric-based learning models.

While FSL has primarily been studied for classification due to its relatively simpler nature, some works have explored its application to semantic segmentation. For example, Prototype Alignment Net (PANet) \cite{Wang2019PANet:Alignment} extended the prototyping networks to segmentation by using semantic masks to calculate the average feature representation of the support set classes. 

Unlike the aforementioned methods, we address point cloud segmentation rather than image classification or segmentation. Still, our approach draws significant inspiration from earlier works. In particular, \cite{Chen2019AClassification} partially inspired our fine-tuning stage, which uses a small multi-layer perceptron (MLP) for final adaptation, and \cite{Chen2019Self-SupervisedClassification} motivated us to pre-train on multiple external dataset in the hopes of improving our prediction accuracy.

\subsection{Few-Shot Learning on Point Clouds}
Research into FSL for point clouds has been far more limited than for images. Most prior works focused on classification of single object point clouds, often derived from 3D models rather than real scans with sensors. For example, Geometric Prototypical Network (GPr-Net) \cite{Anvekar2023GPr-Net:Learning} computed hand-crafted point features and then used a small embedding network to generate prototypes, which were used in combination with hyperbolic distance to perform FSL on ModelNet40 \cite{Wu20153dShapes}. \cite{Ye2023AClassification} attempted to adapt various image FSL algorithms for point cloud classification. These works are difficult to adapt for segmentation. 

A few works have explored point cloud segmentation using metric-based learning networks. Unified 3D Segmenter \cite{Qin2023UnifiedClassifiers} and ProtoSeg \cite{Royen2024ProtoSeg:Method} explored applying prototypical networks to perform semantic, instance, and panoptic segmentation of point clouds, but did not explore few-shot learning, training on limited scenes, or how these networks can be used to improve network generalization across domains. \cite{Zhao2021Few-shotSegmentation} explored FSL for semantic segmentation of point clouds, but their approach relied on constructing expensive k-Nearest Neighbor (k-NN) trees which can be computationally expensive for real-time applications involving large point clouds. Additionally, they only demonstrate 3-way segmentation. In contrast, our method is evaluated for 8-way segmentation and does not require episodic training, which would be difficult to perform with urban public datasets. 

\subsection{Synergistic Multi-Dataset Training}
Multiple FSL studies have shown that larger embedding models trained on more diverse and extensive datasets tend to generalize better to unseen classes and significant domain shifts \cite{Chen2019Self-SupervisedClassification, Hu2022PushingDifference, Ye2023AClassification}. In support of providing more diverse training data for our embedding model, we explore synergistic multi-dataset (SMD) training. SMD training involves joint training a shared model on multiple datasets at once. While simple in theory, differences in data distributions can cause joint training to perform worse than training on a single dataset. To address this challenge, Point Prompt Training \cite{Wu2024TowardsTraining} introduced a modification to normalization blocks that promotes greater consistency when training on multiple datasets, thereby boosting overall performance. PPT has been shown to be effective for transformer-based semantic segmentation models \cite{Wu2024PointStronger} and we explore incorporating it into our projection-based CNN model. 

\subsection{Regularization}
Most transfer learning methods rely on simple classifiers like linear layers to prevent overfitting to small support sets. We argue, however, that this reliance restricts the expressiveness of the feature embeddings, as linear layers force all samples of a class into a compact, linearly separable region of the embedding space. We therefore explore using a lightweight non-linear MLP instead to allow for more complex mappings. To prevent overfitting, we look towards regularizers to smooth the transition boundaries within the embedding space. In particular, we explore the application of Manifold Mixup \cite{Verma2019ManifoldStates}, which has been shown to generate smoother transition boundaries and encourage broader regions of low-confidence predictions by interpolating hidden states during training. This technique is theorized to improve generalization. Prior works have explored using MM for regularizing FSL for image classification \cite{Mangla2019ChartingLearning}. Other works, such as PointMixup \cite{Chen2020PointMixup:Clouds} and SimpliMix \cite{Yang2024SimpliMix:Classification}, proposed input interpolation and point mixing strategies, which can be viewed as adaptations of MM for single-object point clouds. However, these adaptations were designed for single-object point cloud classification and are not easily transferable to segmentation on large-scale scenes. Unlike these prior works, we look to apply MM for point cloud segmentation, incorporating it into our fine-tuning stage. 

\section{Methodology}
The objective of this case study was to design and implement a robust 8-class point cloud segmentation model capable of segmenting individual input scans into the following semantic categories: road, ground, vegetation, people, vehicle, structure, object, and outlier. To support model development and evaluation, a dataset comprising 50 LiDAR scans of the target environment was curated and manually annotated. These scans were selectively sampled from long sequences captured using a vehicle-mounted Ouster OS-1 LiDAR sensor in varied forested and rural environments. The annotated dataset was partitioned into 37 scans for training and 13 for validation. For the remainder of this work, we refer to this dataset as the Target dataset.

As previously stated, our objective can be formulated as an FSL problem. However, in contrast to traditional FSL classification tasks, where the support set typically includes a small number (1–10) of example point clouds, each producing a single prediction, our support set comprises 37 scans, each containing tens of thousands of points. This formulation results in hundreds of thousands of point-level predictions. Consequently, our segmentation task can be viewed as a classification problem with a vast number of simple training samples, inherently mitigating overfitting concerns. As a result, we hypothesize that a trained predictor head consisting of an MLP would perform better than a simple linear or metric-based classifier. Thus, we took inspiration from \cite{Chen2019AClassification} to design a transfer learning approach with a more sophisticated final classifier. 

\begin{figure}[ht]
    \centering
    \begin{subfigure}{0.45\textwidth}
        \centering
        \includegraphics[width=\linewidth]{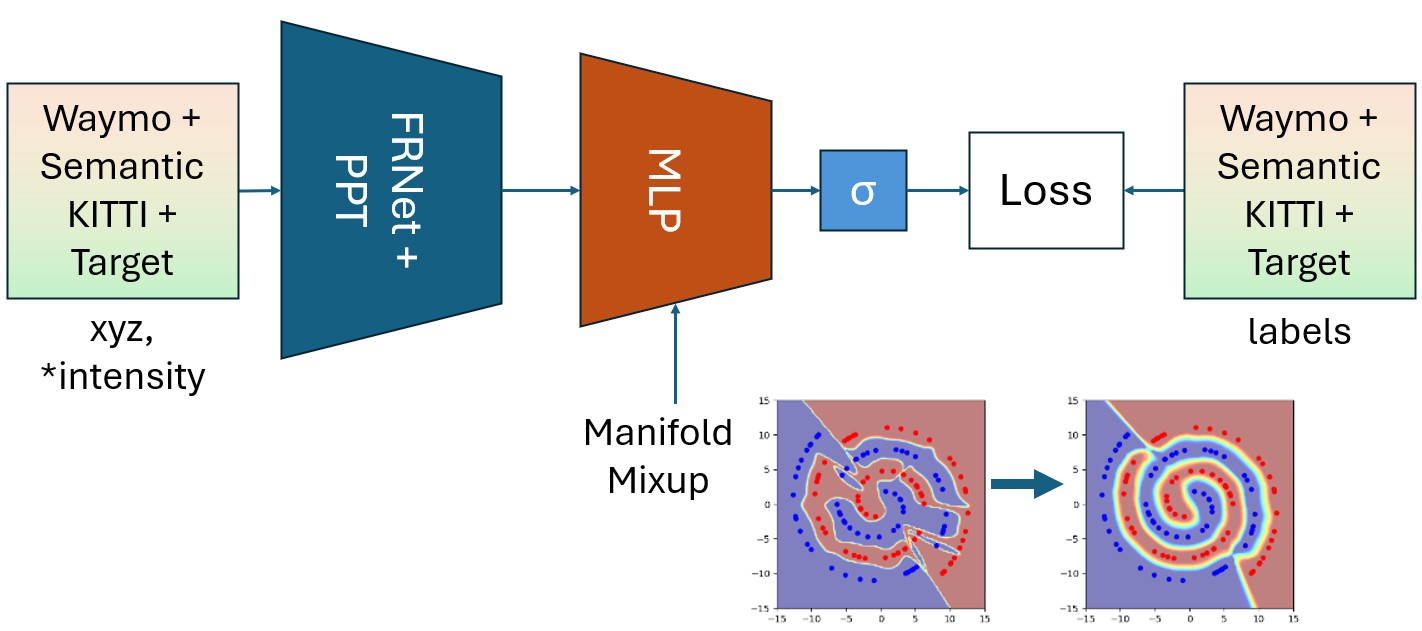}
        \caption{Pretraining Phase}
        \label{fig:method_part1}
    \end{subfigure}
    \hfill
    \begin{subfigure}{0.45\textwidth}
        \centering
        \includegraphics[width=\linewidth]{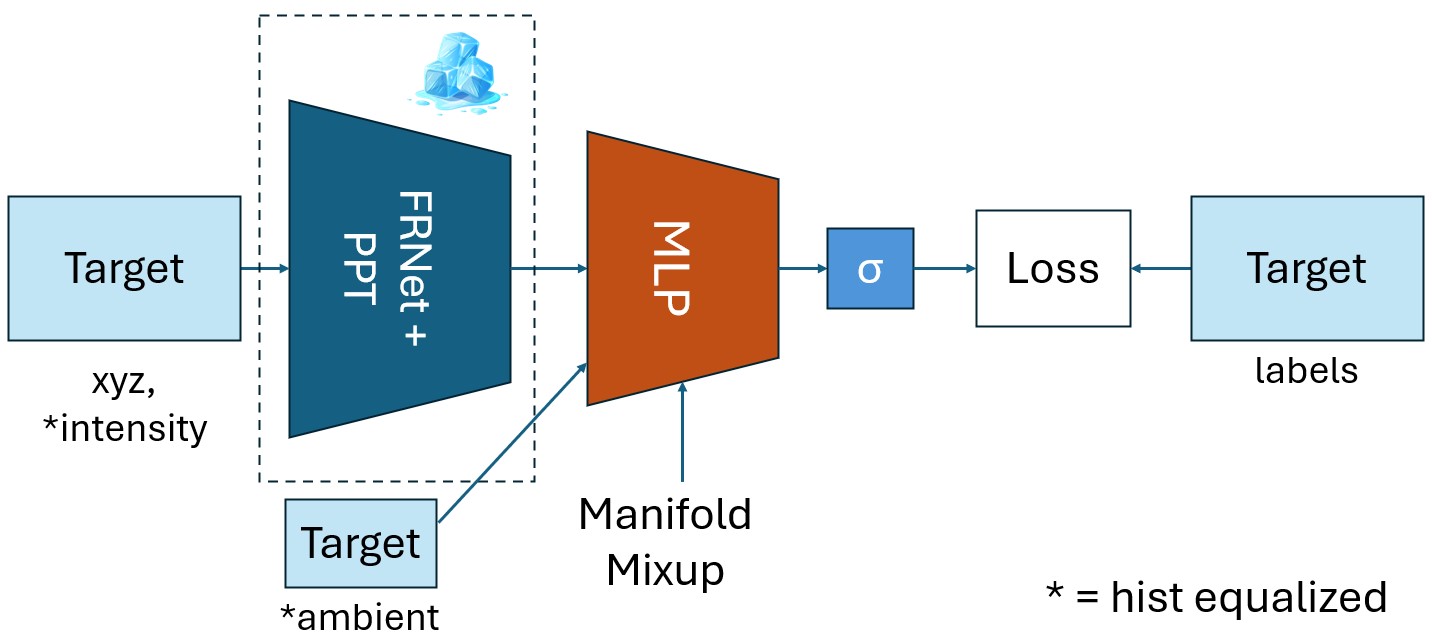}
        \caption{Fine-Tuning Phase}
        \label{fig:method_part2}
    \end{subfigure}
    \caption{An overview of the proposed training methodology for data-efficient training of a semantic segmentation pipeline. Note that the pre-training phase is trained on a mixture of multiple datasets, while the fine-tuning phase only trains on our 37 Target training samples. Image Credit: Manifold Mixup from \cite{Verma2019ManifoldStates}.}
    \label{fig:methodology_overview}
\end{figure}

First, we pre-trained a large feature extractor on a mixture of datasets including Semantic KITTI, Waymo Open Dataset, and the Target dataset. Due to the large number of point clouds available in Semantic KITTI and Waymo Open Dataset, only a few epochs were needed to train a strong feature extractor, reducing the chances of overfitting. We then fixed the feature extractor and trained it on just the Target dataset, fine-tuning the small point-wise multi-layer perceptron (MLP) to assign embedded features to their final class assignments (see Figure \ref{fig:methodology_overview}). Beyond demonstrating robust segmentation performance with limited training data, we further investigated the integration of PPT and MM into a projection-based segmentation framework to enhance model generalization. Finally, we explored the effects of providing additional ambient information to the final classifier to boost segmentation accuracy on the target dataset.

\subsection{Data Preparation}

In order to standardize the inputs across datasets, all datasets were converted to Semantic KITTI format. Points were provided to the model as an N-by-C tensor with N being a variable number of points and C being the number of features per point (C = 4 in our case: $x, y, z, intensity$). Labels were provided point-wise, and the original 34 classes of Semantic KITTI and 23 classes of Waymo Open Dataset were mapped to the target 8 classes of our in-domain dataset. In addition, Waymo provided point clouds from one medium-range overhead LiDAR and four near-field corner LiDARs to provide better coverage of the scene. However, to simplify projection within the model and ensure consistent formatting across datasets, only the overhead LiDAR was used. For our small Target dataset, an additional ambient field was extracted to be used with the final classifier (zeroes used for Semantic KITTI and Waymo Open Dataset). 

To mitigate overfitting, we implemented a diverse set of data augmentation strategies. In addition to the augmentation techniques originally proposed in FRNet \cite{Xu2025FRNet:Segmentation}, we added intensity and ambient dropout, randomized histogram equalization, and minor road/ground-specific rotations.

We applied intensity and ambient dropout with a 20\% probability, inspired by PointNeXt \cite{Qian2022PointNeXt:Strategies}, which demonstrated its effectiveness for indoor segmentation tasks by encouraging models to rely more on geometric structure rather than raw scalar inputs. To further improve robustness to noise and sensor variability, we employed randomized histogram equalization to introduce mild perturbations to the ambient and intensity channels while normalizing their distributions. During training, histogram normalization cutoffs were randomly sampled from narrow ranges (0–5\% for the lower bound, 92–97\% for the upper bound), injecting controlled variability in input features. For evaluation, fixed cutoffs of 2\% and 95\% were used to maintain consistency. Beyond improving generalization, this normalization step ensured consistent feature scaling across datasets and mitigated intensity degradation at longer ranges, enhancing the model's translational invariance. Additionally, we introduced a road/ground-specific rotation augmentation by applying small random yaw rotations (uniformly sampled from ±$1.5^\circ$) to points classified as road or ground in the ground truth, thereby increasing geometric diversity.

\subsection{Model Design}

For our feature extractor module, we chose to use FRNet \cite{Xu2025FRNet:Segmentation}, which demonstrated strong performance in semantic segmentation on the Semantic KITTI dataset. FRNet was chosen due to its fast inference speed, competitive segmentation results, and fast training speed, which measured under 5 hours when training from scratch on Semantic KITTI on an RTX 4090. For semantic segmentation, an inverted-bottleneck MLP (inspired by PointNeXt \cite{Qian2022PointNeXt:Strategies}) followed by a point-wise softmax was used to enable non-linear mappings of FRNet-generated feature embeddings to class labels (see Figure \ref{fig:method_part1}). In prior works that focused primarily on classification, a single linear layer was typically used to perform this mapping due to fears of overfitting to the small in-domain training dataset \cite{Chen2019AClassification, Tian2020RethinkingNeed}. However, we hypothesize that greater modeling capacity was required of this classification stage due to the local nature of the semantic segmentation task. Different local patches of the same object could appear significantly different, and therefore it was not unexpected that they might result in differing feature embeddings. Forcing the embedding space to linearly subdivide into class archetypes may be overly restrictive for semantic segmentation. This issue had been observed in prototypical networks such as \cite{Royen2024ProtoSeg:Method}, which found that allowing multiple prototypes for each class can boost segmentation performance. Additionally, unlike in classification, where only a single feature vector is produced from a point cloud, semantic segmentation generates a feature vector for each point, leading to significantly more classification instances and helping to partially alleviate overfitting concerns.

\subsection{Manifold Mixup}

Although the segmentation task inherently reduces overfitting due to the large number of point-level training samples, the Target training dataset remains limited to just 37 labeled scans. Moreover, the final MLP classifier may learn a fragmented mapping from feature embeddings to class predictions, resulting in sharp decision boundaries that hinder generalization. To mitigate this issue, we explored the use of MM as a regularization technique to promote smoother decision boundaries and improve generalization.

\begin{figure}[htbp]
    \centering
    \includegraphics[width=0.38\textwidth]{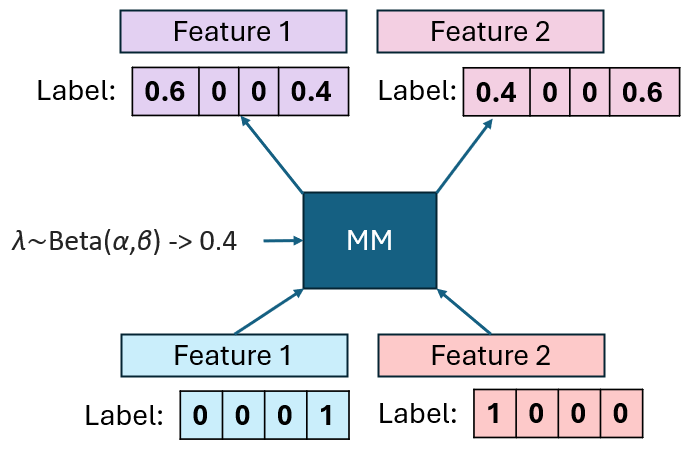}
    \caption{Example application of Manifold Mixup. Two feature vectors (Feature 1 and Feature 2) and their corresponding one-hot labels are selected. A mixing coefficient $\lambda$ is sampled from a Beta distribution (e.g., $\lambda = 0.4$), and the features and labels are linearly interpolated. The resulting mixed feature vector and soft label are passed through the remainder of the model, encouraging the model to produce smooth transitions between classes in the feature space.}
    \label{fig:manifold_mixup}
\end{figure}

MM operates by interpolating both the hidden representations (feature vectors) and their corresponding labels during training. Specifically, given two feature vectors from different points, the method samples a random mixing coefficient $\lambda \in [0,1]$ and generates a new interpolated feature as $\tilde{h} = \lambda h_1 + (1 - \lambda) h_2$, where $h_1$ and $h_2$ are the original features. The corresponding target label is similarly interpolated as $\tilde{y} = \lambda y_1 + (1 - \lambda) y_2$. These mixed features are passed through the remaining layers of the network in place of the originals, and the model is trained to predict the interpolated label (see Figure \ref{fig:manifold_mixup}).

This regularization encourages the model to learn smoother transitions between classes in the embedding space by enforcing that interpolated features map to interpolated labels. Prior work in image classification has shown that this regularization strategy can improve generalization and reduce overfitting, particularly in low-data regimes. We experiment with applying MM within the final MLP prediction head, as it is the most susceptible to overfitting. 

\subsection{Synergistic Multi-Dataset Training}

Prior studies such as \cite{Hu2022PushingDifference} and \cite{Chen2019Self-SupervisedClassification} have shown that incorporating external data into the training pipeline can significantly enhance generalization and improve FSL performance in image classification, even when the external data originates from outside the target domain. Motivated by these findings, we investigate the effects of pretraining on multiple large-scale public datasets, specifically SemanticKITTI and Waymo Open Dataset. 

However, as noted in PPT, training on multiple datasets does not guarantee improved performance. Distributional shifts between datasets can destabilize training and lead to performance degradation across domains. To mitigate this issue, we explore the integration of prompt-normalization into FRNet, which allows for dataset-specific normalization while preserving a shared feature representation across datasets.

\begin{figure}[htbp]
    \centering
    \includegraphics[width=0.45\textwidth]{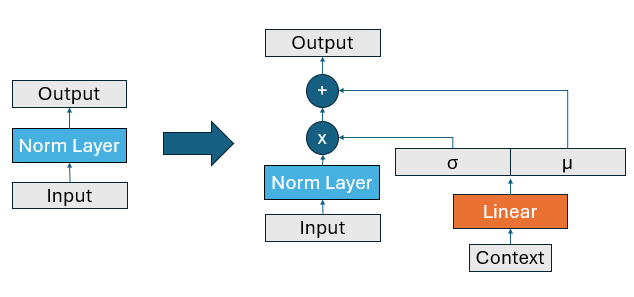}
    \caption{Diagram showing how a normalization layer (e.g. Batch Norm) can be adapted into a prompt-normalization layer. The context embedding is shared across prompt-normalization layers, specific to each dataset, and is learned for each dataset.}
    \label{fig:prompt-norm}
\end{figure}

Prompt-normalization augments all normalization layers by introducing an additional module that generates a mean shift and scaling vector from a learned context embedding. This context embedding is dataset-specific, allowing the model to adaptively adjust feature distributions. The scaling vector is applied element-wise to the output of the original normalization layer, modulating variance, while the mean shift vector introduces a bias correction (see Figure \ref{fig:prompt-norm}). According to the original authors, this mechanism enables prompt-normalization to align feature distributions across datasets, mitigating domain discrepancies.

During fine-tuning, we froze the feature extractor and the linear layers within the prompt-normalization modules while allowing the context embedding to adapt to the new dataset, ensuring effective transfer learning without disrupting the learned feature space.

\subsection{Handling Ambient Values}

During the manual annotation process of the Target dataset, it was noted that the ambient return values, captured by the OS-1 LiDAR, were particularly effective in delineating road boundaries. This observation motivated us to provide ambients as an additional feature to the final model. Since neither Semantic KITTI nor Waymo Open Dataset contains ambient values, we injected histogram-normalized ambients into the prediction head MLP instead of the feature embedding module. The integration was implemented by passing each point’s normalized ambient value through a shared linear layer and concatenating it with the feature embedding before feeding it into the MLP. For points originating from datasets lacking ambient measurements, such as SemanticKITTI and Waymo, the ambient feature was set to zero. 

\section{Experiments}
During pre-training, the proposed framework was trained for 100,000 steps using the AdamW optimizer with a maximum learning rate of 0.002, scheduled via a OneCycleLR learning rate policy. The primary loss function was cross-entropy, supplemented by auxiliary losses as specified in the original FRNet implementation. For fine-tuning, the feature embedding network (FRNet) was frozen, and the remaining model parameters were optimized over 7,600 steps using a maximum learning rate of 0.001, while retaining the same optimizer and scheduling strategy. Throughout all experiments, the validation set was used to monitor model performance and to enable early stopping, thereby mitigating the risk of overfitting. A series of ablation studies were conducted to evaluate the impact of multi-dataset pretraining, PPT, MM, and the incorporation of ambient information within the final MLP classification head.

\subsection{Multi-Dataset Training}

Our experiments indicate that pre-training on multiple datasets is key to improving generalization and enabling robust segmentation with limited in-domain data. Table \ref{tab:dataset_results} reports the mean Intersection‑over‑Union (mIoU) and overall accuracy (Acc) on the Target validation set before and after fine-tuning. Note that for the results in Table \ref{tab:dataset_results}, MM was disabled and PPT was enabled. 

When trained exclusively on the 37 Target scans, the model quickly overfitted: training loss steadily decreased while validation loss stagnated and eventually rose, resulting in poor mIoU and accuracy on the validation set. Training solely on Semantic KITTI or the Waymo Open Dataset without fine-tuning led to even worse performance on the Target domain, with mIoU dropping below 1\%. This severe degradation was likely due to substantial domain shifts and an uninitialized context vector, which caused the model to misclassify points by overpredicting rare classes. Notably, the same models that performed poorly on the Target dataset achieved strong performance on their source domains. The model trained on Semantic KITTI achieved 79.52\% mIoU and 92.86\% overall accuracy on the Semantic KITTI validation split, while the model trained on Waymo yielded 78.55\% mIoU and 92.22\% Acc on its validation set (not shown in Tables). These results highlight that the problem lies in cross-domain generalization rather than in the model’s capacity to learn from data within a single domain.

\begin{table}[htb]
\caption{Validation performance on the Target dataset for models pre‑trained on various datasets. Metrics are shown prior to fine‑tuning (\xmark) and after fine‑tuning (\cmark), as detailed in the methodology section. Relative changes with respect to the Target‑only baseline are given in parentheses (dark green for improvements, dark red for declines).}
\begin{center}
\begin{tabular}{c|c|cc}
\textbf{Pre-training Dataset} & \textbf{Fine-Tuned} & \textbf{mIoU (\%)}      & \textbf{Acc (\%)}        \\
\hline
Target Only & \xmark & 33.51 & 85.52 \\
KITTI Only & \xmark & 0.85 \textcolor{darkred}{(-32.66)} & 6.77 \textcolor{darkred}{(-78.75)} \\
Waymo Only & \xmark & 0.58 \textcolor{darkred}{(-32.93)} & 1.05 \textcolor{darkred}{(-84.47)} \\
Combined  & \xmark & 42.48 \textcolor{darkgreen}{(+8.97)} & 84.53 \textcolor{darkred}{(-0.99)}  \\
\hline
KITTI Only & \cmark & 39.13 \textcolor{darkgreen}{(+5.62)} & 82.46 \textcolor{darkred}{(-3.06)} \\
Waymo Only & \cmark & 36.22 \textcolor{darkgreen}{(+2.71)} & 83.34 \textcolor{darkred}{(-2.18)} \\
Combined & \cmark & 51.80 \textcolor{darkgreen}{(+18.29)}  & 90.76 \textcolor{darkgreen}{(+5.24)} \\
\end{tabular}
\label{tab:dataset_results}
\end{center}
\end{table}

Fine-tuning on the Target dataset after pretraining on Semantic KITTI yielded a modest 5.62\% boost in mIoU, but also a 3.06\% drop in overall accuracy compared to Target-only training. This pattern suggests improved performance on rarer classes, where overfitting is typically most severe, but a slight degradation on the most common classes, which benefit less from more samples and are equally affected by domain shifts. The fine-tuned model pretrained on the Waymo Open Dataset exhibited similar trends, yielding a 2.71\% boost in mIoU and a 2.18\% drop in overall accuracy.

In contrast, pretraining on the combined corpus of the Waymo Open Dataset, SemanticKITTI, and the Target dataset yielded substantial performance improvements even before fine-tuning. Given that the Target data constituted less than 1\% of the mixed dataset, this gain stemmed predominantly from improved generalization. As confirmed by later experiments, this advantage persists regardless of whether PPT is applied. We hypothesize that exposure to diverse sensor characteristics and scene geometries forced the network to learn more intrinsic, geometry-based class representations rather than relying on sensor-dependent cues such as point density or location within the image, thereby improving transfer to the Target data. In comparison, pretraining on a single dataset did not provide sufficient variability to disentangle geometry from sensor and dataset dependent cues, resulting in poorer transfer performance prior to fine-tuning.

Finally, note that the fine-tuning phase boosted performance regardless of the pretraining dataset, demonstrating the efficacy of our two-stage approach. Overall, multi-dataset pretraining followed by Target fine-tuning enabled the model to achieve an mIoU of 51.80\% and an overall accuracy of 90.76\%, corresponding to improvements of 18.29\% in mIoU and 5.24\% in overall accuracy compared to training solely on the Target dataset.

\subsection{Point Prompt Training}

To evaluate the effects of PPT, we trained and fine-tuned a network with and without prompt-normalization. For this ablation study, MM and ambients were kept enabled for all experiments. Table \ref{tab:ppt_results} shows the results of our ablation study. Similarly to \cite{Wu2024TowardsTraining}, we found the effects of PPT to be predominantly positive. Without fine-tuning, adding PPT boosted overall accuracy, but hurt mIoU, suggesting it positively impacted the most common classes, but negatively impacted rarer classes. With fine-tuning, the impact of PPT was entirely positive. Our hypothesis is that without fine-tuning, the Target context vector is not fully optimized, due to the very limited Target data within the mixed dataset. The additional fine-tuning helps to refine the context vector to better capture the differences between datasets, boosting the evaluation results.

Overall, we found that PPT is effective when used with fine-tuning, so we kept it in the final model.

\begin{table}[htb]
\caption{Ablation study on the Target dataset showing validation performance with (\cmark) and without (\xmark) PPT and fine-tuning. Relative changes from the baseline (no PPT, no fine-tuning) are shown in parentheses (dark green: improvements, dark red: declines).}
\begin{center}
\begin{tabular}{c|c|cc}
\textbf{PPT Enabled} & \textbf{Fine-Tuned} & \textbf{mIoU (\%)}      & \textbf{Acc (\%)}        \\
\hline
\xmark & \xmark & 45.40 & 79.89 \\
\cmark & \xmark & 43.44 \textcolor{darkred}{(-1.96)} & 85.62 \textcolor{darkgreen}{(+5.73)} \\
\xmark & \cmark & 46.01 \textcolor{darkgreen}{(+0.61)} & 81.44 \textcolor{darkgreen}{(+1.55)} \\
\cmark  & \cmark & 46.78 \textcolor{darkgreen}{(+1.38)} & 84.91 \textcolor{darkgreen}{(+5.02)}  \\
\end{tabular}
\label{tab:ppt_results}
\end{center}
\end{table}

\subsection{Manifold Mixup}

To quantify the impact of MM, we conducted an ablation study in which models were both pre‑trained and fine‑tuned with and without MM (Table \ref{tab:mm_results}). For these experiments, both PPT and ambients were enabled. During pre‑training, enabling MM improved validation mIoU by 0.96\% and overall accuracy by 1.09\%, suggesting MM can improve generalization on rarely seen domains. However, after fine-tuning, MM led to a worse overall performance than fine-tuning without MM (46.78\% vs 51.80\% mIoU, 84.91\% vs 90.76\% Acc). More testing is needed to better understand this limitation, but we hypothesize that the generalization gains made by MM may have come at the expense of specificity on in-domain datasets. Thus, while MM boosted performance on rarely seen domains, leading to the initial gains, it overly constrained the capacity of the final MLP, limiting its ability to map the frozen out-of-domain feature embeddings to the correct classes.

Overall, we achieved best performance by disabling MM and finetuning on the Target dataset. Thus this configuration was chosen for the final model.

\begin{table}[htb]
\caption{Ablation study on the Target dataset showing validation performance with (\cmark) and without (\xmark) MM and fine-tuning. Relative changes from the baseline (no MM, no fine-tuning) are shown in parentheses (dark green: improvements, dark red: declines).}
\begin{center}
\begin{tabular}{c|c|cc}
\textbf{MM Enabled} & \textbf{Fine-Tuned} & \textbf{mIoU (\%)}      & \textbf{Acc (\%)}        \\
\hline
\xmark & \xmark & 42.48 & 84.53 \\
\cmark & \xmark & 43.44 \textcolor{darkgreen}{(+0.96)} & 85.62 \textcolor{darkgreen}{(+1.09)} \\
\xmark & \cmark & 51.80 \textcolor{darkgreen}{(+9.32)} & 90.76 \textcolor{darkgreen}{(+6.23)} \\
\cmark  & \cmark & 46.78 \textcolor{darkgreen}{(+4.30)} & 84.91 \textcolor{darkgreen}{(+0.38)}  \\
\end{tabular}
\label{tab:mm_results}
\end{center}
\end{table}

\subsection{Ambient Inclusion}

To evaluate the impact of incorporating ambient information into the final prediction head, we conducted an ablation study comparing model performance with and without ambient values. For this experiment, MM was disabled and PPT was enabled, as this configuration had been identified as optimal in earlier ablation studies. The results, shown in Table \ref{tab:ambient_results}, indicate that including ambient information generally improved mIoU, though it caused a slight decrease in overall accuracy.

Before fine-tuning, incorporating ambient values increased mIoU by 8.17\% but reduced overall accuracy by 1.66\%. The largest gains were observed in the vehicle, structure, and outlier classes, while performance for ground and road classification declined. After fine-tuning, the inclusion of ambient information led to a smaller 3.57\% improvement in mIoU and a negligible 0.01\% decrease in overall accuracy compared to fine-tuning without ambient values. In this case, the road, structure, and object classes saw the most significant improvements in class-wise mIoU, while ground and vegetation classes experienced less than a 0.5\% decrease.

We hypothesize that prior to fine-tuning, the model only leveraged ambient cues for coarse categorization, as much of the training dataset lacked ambient values. Such reliance likely favored classes with extreme ambient characteristics (e.g., outliers or vehicle headlights). Post fine-tuning, the network appeared to incorporate ambient information in a more fine-grained manner, improving road segmentation and helping to distinguish between visually similar classes such as signs and structures, which often co-occur but differ significantly in their ambient signatures.

Given the performance gains observed during fine-tuning, particularly for road segmentation, ambient information was retained in the final model configuration.

\begin{table}[htb]
\caption{Ablation study on the Target dataset showing validation performance with (\cmark) and without (\xmark) ambient features and fine-tuning. Relative changes from the baseline (no ambient, no fine-tuning) are shown in parentheses (dark green: improvements, dark red: declines).}
\begin{center}
\begin{tabular}{c|c|cc}
\textbf{Uses Ambient} & \textbf{Fine-Tuned} & \textbf{mIoU (\%)}      & \textbf{Acc (\%)}        \\
\hline
\xmark & \xmark & 34.31 & 86.19 \\
\cmark & \xmark & 42.48 \textcolor{darkgreen}{(+8.17)} & 84.53 \textcolor{darkred}{(-1.66)} \\
\xmark & \cmark & 48.23 \textcolor{darkgreen}{(+13.92)} & 90.77 \textcolor{darkgreen}{(+4.58)} \\
\cmark  & \cmark & 51.80 \textcolor{darkgreen}{(+17.49)} & 90.76 \textcolor{darkgreen}{(+4.57)}  \\
\end{tabular}
\label{tab:ambient_results}
\end{center}
\end{table}

\subsection{Final Model and Qualitative Results}

Given the results of the ablation studies, our final model is pre-trained on a mixed dataset of Waymo Open Dataset, Semantic KITTI, and our custom Target dataset. The model uses PPT, but does not perform MM. We also include ambient in the final MLP prediction head. The model is fine-tuned using only our Target dataset. The final results are shown in Table \ref{tab:final_results}.

It is worth noting that the Target validation set includes only a single distant person, resulting in just 14 labeled points for the ``people'' class. Future work may consider expanding the dataset to include more scenes featuring people, thereby enabling a more meaningful evaluation of performance on this class. Additionally, the ``outlier'' class presents a unique challenge, as it is not labeled in the Waymo Open Dataset and is labeled much more aggressively in Semantic KITTI, resulting in inconsistent supervision and reduced reliability for this class.

\begin{table}[htbp]
\caption{Final model performance on the Target validation set measured as mean Intersection-over-Union (mIoU), overall accuracy (Acc), mean class accuracy (mAcc), and per-class IoU results.}
\begin{center}
\renewcommand{\arraystretch}{1.5}
\resizebox{0.5\textwidth}{!}{
\begin{tabular}{|c|c|c|c|c|c|c|c|c|c|c|}
\hline
\multirow{2}{*}{\rotatebox{90}{\textbf{mIoU (\%)   }}} & 
\multirow{2}{*}{\rotatebox{90}{\textbf{Acc (\%)    }}} & 
\multirow{2}{*}{\rotatebox{90}{\textbf{mAcc (\%)   }}} & 
\multicolumn{8}{c|}{\textbf{Per-Class IoU (\%)}} \\
\cline{4-11}
& & & 
\rotatebox{90}{Ground} & 
\rotatebox{90}{Road} & 
\rotatebox{90}{Vegetation } & 
\rotatebox{90}{Structure} & 
\rotatebox{90}{Vehicle} & 
\rotatebox{90}{People} & 
\rotatebox{90}{Object} & 
\rotatebox{90}{Outliers} \\
\hline
51.80 & 90.76 & 60.45 & 74.02 & 59.49 & 93.06 & 53.96 & 71.93 & 0.0 & 40.97 & 20.94 \\
\hline
\end{tabular}
}
\label{tab:final_results}
\end{center}
\end{table}

\begin{figure}[ht]
    \centering
    \begin{subfigure}{0.47\textwidth}
        \centering
        \includegraphics[width=\linewidth]{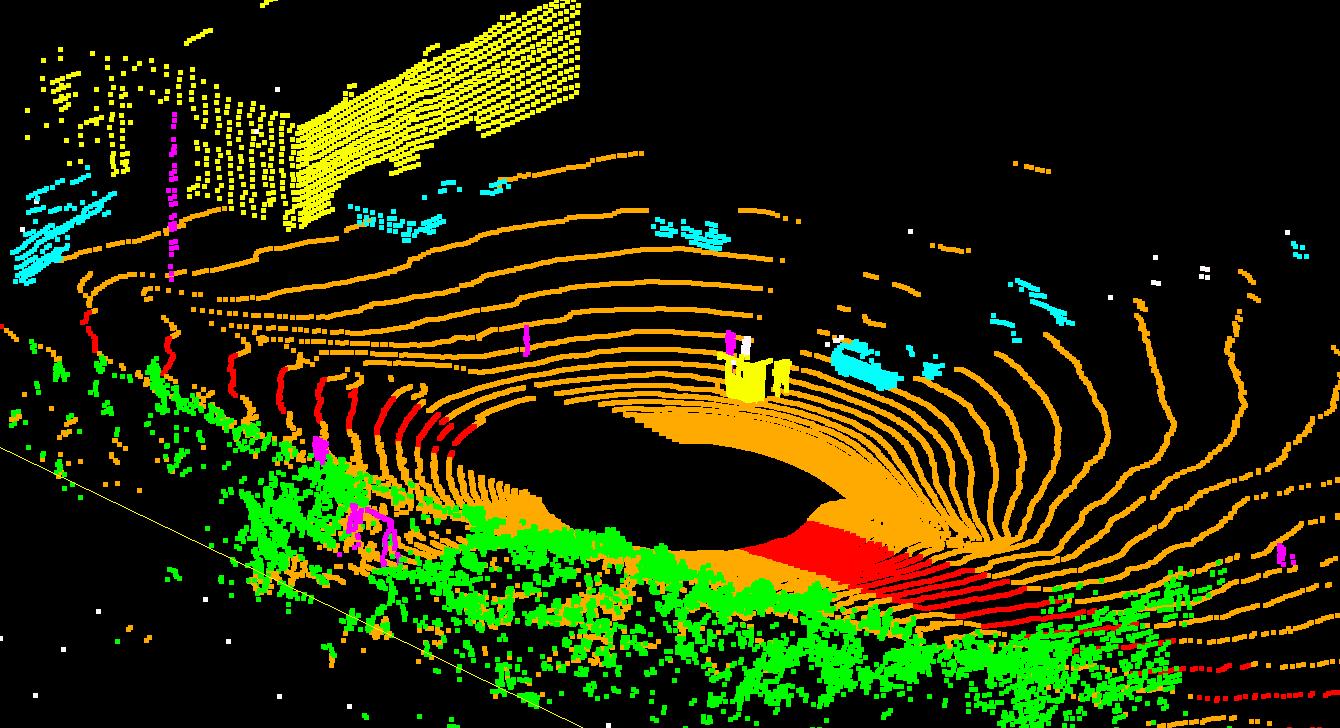}
        \caption{Ground Truth Labels}
        \label{fig:ground_truth}
    \end{subfigure}
    \hfill
    \begin{subfigure}{0.47\textwidth}
        \centering
        \includegraphics[width=\linewidth]{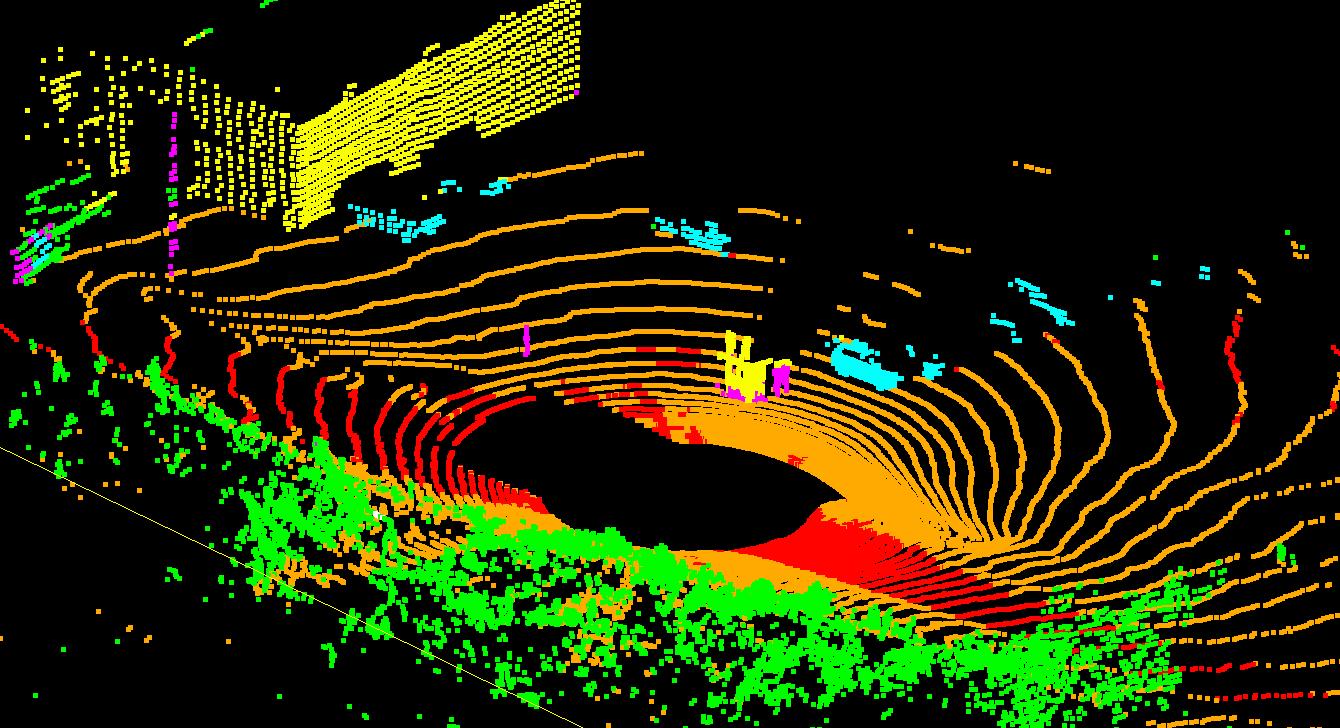}
        \caption{Predicted Labels}
        \label{fig:predictions}
    \end{subfigure}
    \caption{An example of the predicted labels for a diverse scene within the Target validation set. The colors correspond to class labels as follows: red is road, orange is ground, green is vegetation, yellow is structure, cyan is vehicle, magenta is object, and white is outlier. This scene does not contain any points labeled as people.}
    \label{fig:segmented_example}
\end{figure}

An illustrative segmentation result from the Target validation set is shown in Figure \ref{fig:segmented_example}. Despite the subtle boundaries of the dirt road, which were fairly challenging even for human annotators to delineate, the model produced reasonable predictions. From this scene, we can see that dominant classes like road, ground, and vegetation are extracted fairly accurately. Extraction of the "object" class is less reliable, with several objects mislabeled as vegetation or structure. Overall, the model achieves notable results given the limited in-domain dataset.

\section{Conclusion}
In this case study, we propose a data-efficient model and training framework for effective semantic segmentation of point clouds with limited annotated in-domain data. Through experiments, we demonstrated the effectiveness of our transfer learning approach, boosting prediction mIoU by 18.29\% and overall accuracy by 5.24\%. Our experiments demonstrated that pre-training on multiple datasets was key to improving cross-dataset generalization and enabling effective transfer learning. We also presented comprehensive ablation studies, informing our usage of Point Prompt Training and ambient features within our network and evaluating the effectiveness of Manifold Mixup on our approach.

While our approach yields substantial performance improvements, several promising avenues remain for future exploration to further enhance the model’s effectiveness. Several data augmentation techniques were added to the original FRNet to increase data variety, but due to time constraints, we were unable to perform ablative studies to validate their effectiveness. Additionally, because ambient values are introduced after feature extraction, the feature extractor itself is unable to leverage this information during pre-training. An alternative approach which uses unlabeled in-domain data to predict ambients for Semantic KITTI and Waymo Open Dataset could be investigated to enable earlier injection of the ambient features. 

Another avenue for further study is to introduce confidence scores into the prediction head. During data annotations, it was noted that even for human annotators, it was difficult to determine certain classes without resorting to external data. These ambiguous points can harm the training by punishing the model for incorrect predictions despite the innate ambiguity, leading to less stable training. Adding confidence scores and incorporating them into the loss function could help distinguish these ambiguous cases, leading to better overall predictions and easier post-processing of the predicted labels. 

Further work will explore these avenues, as well as test the model on edge-computing hardware for real-time applications.

\bibliographystyle{ieeetr}
\bibliography{IEEEabrv, main}
\end{document}